\documentstyle[11pt]{article}

\def \1{\'{\i}}

\def \&{&=&}

\newcommand{\be}{\begin{equation}}

\newcommand{\ee}{\end{equation}}

\newcommand{\beq}{\begin{eqnarray}}

\newcommand{\eeq}{\end{eqnarray}}

\newcommand{\ba}{\begin{array}}

\newcommand{\ea}{\end{array}}

\begin{document}

\setcounter{equation}{0}

\setcounter{section}{0}

\title{\Large \bf Darboux transformations for a Bogoyavlenskii equation in $2+1$ dimensions}

\author{ \bf P.G. Est\'evez\footnote{e-mail: pilar@sonia.usal.es } and
G.A. Hern\'aez   \\ {\small
\bf Area de F\1sica Te\'orica}\\ {\small \bf Facultad de F\1sica}\\
{\small \bf Universidad de Salamanca}\\ {\small \bf 37008 Salamanca.
Spain}\\}

\maketitle

\begin{abstract} 
 We use the singular manifold method to obtain the Lax pair, 
Darboux transformations and soliton solutions for a (2+1) dimensional integrable equation. 
\end{abstract} \vspace*{0.3in}

\section{Introduction}

The equation 
\begin{equation} u_{xt}+{1\over 4}u_{xxxy}+u_xu_{xy}+{1\over2}u_{xx}u_y+
{1\over 4}\int{u_{yyy}}dx=0 \end{equation} 
obtained by Bogoyavlenskii in \cite{b90}, has been recently rederived as a (2+1)
 dimensional reduction of an equation which pretended to be a (3+1) dimensional
  generalization of the potential KP equation \cite{ytf981}, \cite{ytf982},\cite{Z99}.

This equation is integrable in the sense of having the Painlev\'e property \cite{ytf981}.
We can write equation (1) in a more appropiate way as the system
\begin{eqnarray}   \nonumber 0&=&u_y-m_x \\ 
0&=&u_{xt}+{1\over4}u_{xxxy}+u_xu_{xy}+{1\over2}u_{xx}u_y+{1\over4}m_{yy} \end{eqnarray} 

\section{Singular Manifold Method}

\subsection {Leading term analysis}

The Painlev\'e property for this equation means that all solutions of (2) can be expanded as a 
generalized 
Laurent expansion in the neighbourhood of the manifold of movable singularities $\chi(x,y,t)$ 
which is an 
arbitrary function. This expansion should be \cite{wtc83}
\begin{equation}  u(x,y,t)= \sum_{j=0}^{\infty}u_j(x,y,t)\chi(x,y,t)^{j-\alpha}, \, 
m(x,y,t)=\sum_{j=0}^{\infty}m_j(x,y,t)\chi(x,y,t)^{j-\beta} \end{equation}
where the $u_j$ and $m_j$ are analytical functions of $x$, $y$ and $t$ in the neighbourhood of 
$\chi=0$.
The leading term analysis yields:
\begin{equation} \alpha=\beta=1 \quad u_0=2\chi_x \quad m_0=2\chi_y \end{equation}

\subsection {Truncated expansions. B\"{a}cklund transformations}

The singular manifold method requires the truncation of expansions (3) at the constant level 
$j=1$ 
\cite{w83}. This implies that the manifold of movable singularities $\chi$ is no longer an 
arbitrary function 
but it has to satisfy some equations, called the singular manifold equations, that we will see 
later on. Due 
to this fact, we denote this manifold as $\phi$ and call it the {\bf singular manifold} in the 
sense that it 
is singularized by the truncation condition.The truncated solutions are then
\begin{equation}  u'= u+ {{2\phi_x}\over\phi}, \quad m'=m+{{2\phi_y}\over\phi} \end{equation}
Substitution of (5) in equation (2) provides us a polynomial in $\phi$. Setting to zero the 
coefficents of 
this polynomial we find that $u$ and $m$ must be solutions of (2), which means that eqs. (5) 
can be 
considered a {\bf B\"{a}cklund transformation} between two solutions of (2).

\subsection{Expression of the solutions in terms of the Singular Manifold}

From the coefficients of the polynomial in $\phi$ we obtain
\begin{eqnarray}  u_x&=&-{1\over 4}p_x^2+{1\over2}p_y-{1\over2}v_x-{1\over4}v^2+h(y,t),\\ 
u_y&=&-2w-v_y-p_xp_y-2p_xh(y,t)\\ \nonumber m_y&=&-2p_t-p_{xxy}-{1\over 
2}p_{xx}^2+p_x^2p_y-{1\over2}p_y^2-p_{xy}v-p_xp_{xx}v \\ & & \mbox{}-{1\over 
2}p_x^2v^2+2wp_x++2p_x^2h(y,t)-2p_yh(y,t) \end{eqnarray}
where $h(y,t)$ arises from an integration in $x$ and $v$, $w$ and $p$ are defined in terms of 
the singular 
manifold as
\begin{equation} v={\phi_{xx}\over\phi_x}, \quad w={\phi_t\over\phi_x}, \quad 
p_x={\phi_y\over\phi_x} 
\end{equation}
\subsection {Singular Manifold Equations}

From the compatibility conditions between definitions (9) we obtain the following generic 
equations
\begin{eqnarray}  \nonumber \phi_{xxt}=\phi_{txx} &\Longrightarrow& v_t=(w_x+vw)_x \\ 
\nonumber 
\phi_{xxy}=\phi_{yxx} &\Longrightarrow& v_y=(p_{xx}+vp_x)_x \\
  \phi_{yt}=\phi_{ty} &\Longrightarrow& p_{xt}=w_y+wp_{xx}-p_xw_x \end{eqnarray}
Moreover, taking the cross derivatives of (6) and (7)($u_{xy}=u_{yx}$) we have
\begin{eqnarray} \nonumber & &  \left(p_{xxx}+4w+2p_xp_y+p_x\left(v_x-{v^2\over 
2}\right)+4p_xh(y,t)\right)_x+ \\ & & \mbox{} +\left(p_y-{1\over 
2}p_x^2+2h(y,t)\right)_y+p_{xx}\left(v_x-{v^2\over 2}\right)=0 \end{eqnarray}
The set (10-11) constitutes the {\bf Singular Manifold Equations}.
\section{Lax Pair}

\subsection{Painlev\'e Analysis on the singular manifold equations}

The singular manifold equations (10) and (11) can be considered as a system of non-linear 
coupled PDE's in 
$v$, $w$ and $p$ and we can perform the Painlev\'e analysis over them. It's not difficult to 
see, following 
the same procedure as in the previous section, that leading term analysis in the singular 
manifold equations 
yields the truncated expansions: 
\begin{equation}  v={\psi_x^+\over\psi^+}+{\psi_x^-\over\psi^-}, \quad 
p_x={\psi_x^+\over\psi^+}-{\psi_x^-\over\psi^-}, \quad 
p_y={\psi_y^+\over\psi^+}-{\psi_y^-\over\psi^-}, \quad 
p_t={\psi_t^+\over\psi^+}-{\psi_t^-\over\psi^-} \end{equation}
where we have two singular manifolds $\psi^+$ and $\psi^-$ because the Painlev\'e expansion 
has  two 
branches \cite{eg97}, \cite{ecg98}.
Taking the $t$ and $y$ derivatives of eqs. (12) and using (10) to integrate them in $x$ we 
have
\begin{equation}  w_x+vw={\psi_t^+\over\psi^+}+{\psi_t^-\over\psi^-}, \quad 
p_{xx}+vp_x={\psi_y^+\over\psi^+}+{\psi_y^-\over\psi^-} \end{equation}
Moreover, integration of (12) yields
\begin{equation}  \phi_x=\psi^+\psi^-, \quad \phi_y=\psi_x^+\psi^--\psi^+\psi_x^- 
\end{equation}
These equations allow us to obtain the singular manifold $\phi$ from the eigenfunctions 
$\psi^+$ and 
$\psi^-$.

\subsection{Linearization of the singular manifold equations: Lax pair}
From (12) and (13) we can easily obtain, with the help of (6)-(8), the {\bf spatial part of 
the Lax pair}:
\begin{eqnarray} \nonumber \psi_{xx}^+ + (u_x-h)\psi^+-\psi_y^+&=&0\\ 
\psi_{xx}^-+(u_x-h)\psi^-+\psi_y^-&=&0 
\end{eqnarray}
where $h(y,t)$ is the spectral parameter.

The {\bf temporal part of the Lax pair} can be obtained by combining equations (12)-(14) with 
(6)-(8)
in order to eliminate $w$ from them  to obtain:
\begin{eqnarray} \nonumber \psi_t^++{1\over 2}\psi_{yy}^+-{1\over 
4}\left(u_{xy}-m_y-2h_y\right)
\psi^++{1\over 2}u_y\psi_x^++h\psi_y^+&=&0\\ \psi_t^--{1\over 2}\psi_{yy}^--{1\over 
4}\left(u_{xy}+m_y-2h_y\right)\psi^-+{1\over 2}u_y\psi_x^-+h\psi_y^-&=&0 \end{eqnarray}
From the compatibility condition of the equations of the Lax pair among themselves and with 
equation (2) we 
have the condition
\begin{equation} h_t+hh_y=0 \end{equation}
which means that {\bf the problem is  non-isospectral}.
\section{Darboux Transformations}

As far as $u'$ and $\omega'$ are also solutions of (2), we can define a new singular manifold 
$\phi'$ for 
them by means of two eigenfunctions $\psi'^+$ and $\psi'^-$ with eigenvalue $h_2$ as:
\begin{equation}  \phi_x'=\psi'^+\psi'^-, \quad\quad \phi_y'=\psi_x'^+\psi'^--\psi'^+\psi_x'^- 
\end{equation}
where $\psi'^+$ and $\psi'^-$ must satisfy the Lax pair for $u'$ with spectral parameter 
$h_2$. We can now 
consider the Lax pair itself as a system of coupled nonlinear equations in $u'$, $m'$, 
$\psi'^+$ and 
$\psi'^-$ \cite{ks91} and hence we can expand the fields and eigenfunctions in the Painlev\'e 
series
\begin{equation}  u'=u+{{2\phi_{1x}}\over\phi_1}, \quad m'=m+{{2\phi_{1y}}\over\phi_1} 
\end{equation} 
\begin{equation} \psi'^+=\psi_2^+-{{\psi_1^+\Omega^+}\over\phi_1}, \quad  
\psi'^-=\psi_2^--{{\psi_1^-\Omega^-}\over\phi_1} \end{equation} 
\begin{equation} \phi'= \phi_2+{\Delta\over\phi_1} \end{equation}
where $\psi_1^+$ and $\psi_1^-$ are eigenfunctions of $u$ and $m$ with eigenvalue $h_1$ and 
$\psi_2^+$ and 
$\psi_2^-$ are eigenfunctions with eigenvalue $h_2$ satisfying the Lax pair for $u$.

Substitution of the truncated expansions in the Lax pair for $\psi'^+$ and $\psi'^-$ yields 
(we have 
used MapleV to handle the calculation): 
\begin{equation}  \Omega_x^+=\psi_2^+\psi_1^-, \quad  
\Omega_y^+=\Omega^+(h_1-h_2)+\psi_{2x}^+\psi_1^--\psi_2^+\psi_{1x}^- \end{equation}
\begin{equation}  \Omega_x^-=\psi_1^+\psi_2^-, \quad 
\Omega_y^-=\Omega^-(h_2-h_1)+\psi_{1x}^+\psi_2^--\psi_1^+\psi_{2x}^- \end{equation}
and from the substitution of (21) in (18) we have
\begin{equation} \Delta=-\Omega^+\Omega^- \end{equation}
The set of equations (19)-(21) with $\Omega^+$ and $\Omega^-$ given by (22)-(23), constitutes 
a transformation of 
eigenfunctions and potentials that leaves invariant the Lax pair and hence it can be 
considered a {\bf 
Darboux transformation} \cite{ms91}.

\section{Hirota's $\tau$-function}

Hirota's bilinear method is a tool used in many references to obtain multisolitonic solutions 
for nonlinear 
PDE's. N-soliton solutions of (2) have been constructed with this method in \cite{ytf982}. 
In this section we shall build  the $\tau$-functions of Hirota's method through the iteration 
of the singular 
manifold. Using equation (21)
\begin{equation} \phi'=\phi_2-{{\Omega^+\Omega^-}\over \phi_1} \end{equation}
which defines a singular manifold for $m'$, we can use such manifold in order to construct an 
iterated 
solution
\begin{equation} u''= u'+{\phi_x'\over\phi'} \end{equation}
Substituting  equation (5) for $u'$ in (26)  we have:
\begin{equation} u''=u+{{2\tau_x}\over\tau} \end{equation}
where
\begin{equation} \tau=\phi'\phi_1=\phi_1\phi_2-\Omega^+\Omega^- \end{equation}
is Hirota's $\tau$-function \cite{h85}.

\section{Solutions}

In this section we shall obtain the one and two soliton solutions of equation (2) using the 
results of the 
previous sections. We start from the seminal solutions 
\begin{equation} u=0 \quad\quad m=0 \end{equation}
In this case, and restricting ourselves to the case when $h_1$ and $h_2$ are constants,  
non-trivial simple 
solutions of the Lax pair  are:
\begin{equation}  \psi_i^+=\exp\left[\alpha_i^+x+\beta_i^+\left(y-\left({\alpha_i^+}^2-{1\over 
2}\beta_i^+\right)t\right)\right]\end{equation} 
\begin{equation}  \psi_i^-=\exp\left[\alpha_i^-x+\beta_i^-\left(y-\left({\alpha_i^-}^2+{1\over 
2}\beta_i^-\right)t\right)\right]\end{equation} 
where $i=1,2$ and $\alpha_i^+$, $\alpha_i^-$, $\beta_i^+$ and $\beta_i^-$ are constants 
related with the 
spectral parameter and among themselves by
\begin{equation} h_i=\alpha_i^{-2}+\beta_i^-=\alpha_i^{+2}-\beta_i^+ \end{equation}
Integration of (14) yields
\begin{equation} \phi_i={c_i\over{\alpha_i^++\alpha_i^-}}\left(1+F_i\right) \end{equation}
where
\begin{equation} 
F_i=\exp\left\{(\alpha_i^++\alpha_i^-)x+(\beta_i^++\beta_i^-)y-\left(\beta_i^+\alpha_i^{+2}+
\beta_i^-\alpha_i
^{-2}+{\beta_i^{-2}-\beta_i^{+2}\over 2}\right)t\right\} \end{equation}
Integrating (22) and (23) we have:
\begin{equation} \Omega^+={1\over{\alpha_2^++\alpha_1^-}}\left(d^++\psi_2^+\psi_1^-\right), 
\quad  
\Omega^-={1\over{\alpha_1^++\alpha_2^-}}\left(d^-+\psi_1^+\psi_2^-\right) \end{equation}
where $d^+$ and $d^-$ are arbitrary constants.

\noindent $\bullet$ {\bf One soliton solution} 
The first iteration provides the solution
\begin{equation} u_x=2\partial_{xx}[\ln\phi_1] \end{equation}
with $\phi_1$ given by (33).
It corresponds with one line soliton.

\noindent $\bullet$ {\bf Two soliton solution}
From the second iteration  we have:
\begin{equation} u''=2\partial_{xx}[\ln\tau] \end{equation}
with
\begin{equation} 
\tau=\phi_1\phi_2-\Omega^+\Omega^-={{c_1c_2}\over{(\alpha_1^++\alpha_1^-)(\alpha_2^++\alpha_2^
-)}}\left(1+F_1
+F_2+A_{12}F_1F_2\right) \end{equation}
$F_i$ is given by (34) and 
\begin{equation} 
A_{12}={{(\alpha_2^+-\alpha_1^+)(\alpha_2^--\alpha_1^-)}\over{(\alpha_2^++\alpha_1^-)(\alpha_1
^++\alpha_2^-)}
} \end{equation}
Equation (38) represents the interaction of two line solitons.
When $\alpha_2^+=\alpha_1^+$ or $\alpha_2^-=\alpha_1^-$, the interaction term $A_{12}$ 
vanishes and this 
special case is termed the {\bf resonant state} \cite{el95}.

\end{document}